\documentstyle[11pt]{article}

\def\afour{ \setlength{\topmargin}{0mm} \setlength{\headheight}{0mm}
\setlength{\headsep}{0mm} \setlength{\textwidth}{6in}
\setlength{\textheight}{248mm} \setlength{\oddsidemargin}{.25in}
\setlength{\evensidemargin}{.25in} } 

\newcommand\eq[1]{Eq.~(\ref{#1})}

\newcommand{\sub}[1]{_{\mbox{\scriptsize#1}}}
\newcommand{\su}[1]{^{\mbox{\scriptsize#1}}}

\newcommand\ee{\end{equation}} \newcommand\be{\begin{equation}}
\newcommand\eea{\end{eqnarray}} \newcommand\bea{\begin{eqnarray}}

\newcommand\mone{^{-1}} \newcommand\mtwo{^{-2}}
 \newcommand\mfour{^{-4}} \newcommand\mhalf{^{-1/2}}

\newcommand\half{^{1/2}} 
 
\newcommand\quarter{^{1/4}}

\newcommand\lsim{\mathrel{\rlap{\lower4pt\hbox{\hskip1pt$\sim$}}
    \raise1pt\hbox{$<$}}}
    \newcommand\gsim{\mathrel{\rlap{\lower4pt\hbox{\hskip1pt$\sim$}}
    \raise1pt\hbox{$>$}}}

\newcommand\diff{\mbox d}

 \def\calp{{\cal P}} \def\calr{{\cal R}}

\afour

\def\rt{R^{(3)}}  \def\kl{_{kl}} \def\lm{_{l m}} \def\klm{_{kl
m}}  \def\klm{_{kl m}} 
\def\calr{{\cal R}}

\begin{document}

\begin{flushright}
LANC-TH9512   ASTRO-PH/9501113
  \end{flushright}

\begin{center}

{\LARGE \bf THE GRISHCHUK-ZELDOVICH EFFECT}

\vspace{4mm}
\noindent
{\LARGE \bf IN THE OPEN UNIVERSE}

\vspace{.3in} {\large David H.  Lyth}

\vspace{.6 cm} {\normalsize \em School of Physics and Chemistry, \\
University of Lancaster, Lancaster LA1 4YB.~~~U.~K.}\\

\end{center}

\bigskip
\begin{center}
{\large \bf Introduction}
\end{center}

\smallskip
\noindent
When considering perturbations in an open universe,
cosmologists retain only
sub-curvature modes (defined as eigenfunctions of the Laplacian
whose eigenvalue is less than $-1$ in units
of the curvature scale, in contrast with the super-curvature
modes whose eigenvalue is between $-1$ and $0$).
Mathematicians have known for
almost half a century that all modes must be included
to generate the most general {\em
homogeneous Gaussian random field},
despite the fact that any square integrable {\em function}
can be generated using only the sub-curvature modes.
The former mathematical object, not the latter,
is the relevant one for physical applications.
This article summarizes recent work with A. Woszczyna.
The mathematics is briefly explained in a language accessible to
physicists. Then the effect on the cmb of any super-curvature
contribution is considered, which generalizes to $\Omega_0<1$ the
analysis given by Grishchuk and Zeldovich in 1978.

\bigskip
\begin{center}
{\large \bf The mode expansion}
\end{center}

\smallskip
\noindent
According to the Einstein field equation, the
energy density $\Omega$ of the universe
is given by
\be 1-\Omega= -\frac K{(aH)^2} \label{omega} \ee
Here $K$ is a constant, $H=\dot a/a$ is the
Hubble parameter, and $\Omega$ is the energy density measured in units of the
critical density $3H^2/8\pi G$.
The spatial curvature scalar
is $R^{(3)}=6K/a^2 \label{rthree}$ and we set
 $K=-1$ so that $a$ is the curvature scale. Then  the
 case $\Omega=1$ corresponds to the limit $a\to\infty$, with physical
 distances like  $H^{-1}$
 remaining constant.

We are interested in the stochastic properties of the perturbations, at
fixed
time.  To define them we will take the approach of considering an
ensemble of
universes of which ours is supposed to be one.
If, in some region of space,  a perturbation $f$ can be written as
a sum of terms, with the coefficient of each term having
an independent Gaussian
probability distribution, it is a {\em Gaussian random field},
and its stochastic properties are completely determined by its
correlation function. There is no requirement that the
region of space be infinite, or that the
terms be
linearly independent.
If the correlation function depends only on the geodesic distance
between
the points, the field is said to be {\em homogeneous}
with respect to the group of transformations that preserve this
distance.
We assume that each
cosmological perturbation in the observable universe is
a typical realization of some homogeneous Gaussian random field.

Spherical coordinates are defined by the line element
 \be \diff l^2 =a^2[ \diff r^2+\sinh^2  r(\diff\theta^2 +\sin^2\theta
\diff\phi^2)]
 \label{elev} \ee
Any homogeneous Gaussian
random field can be generated${}^{1,2,3}$
by expanding it in eigenfunctions of the
Laplacian with eigenvalue $(k/a)^2<0$,
\be
f(  r,\theta,\phi,t)=\int_0^\infty \diff k \sum\lm \tilde f\klm(t)
Z\klm(  r,\theta,\phi) \label{expansion}
\ee
Here $q^2=k^2-1$, and  the
mode functions are $Z\klm=\Pi_{kl}(  r) Y_{lm}(\theta,\phi)$. For
$q^2>0$ the radial functions are
\bea
\Pi_{kl}&\equiv & N_{kl}\tilde\Pi_{kl}\\
\tilde \Pi_{kl} &\equiv& |q|\mtwo (\sinh
  r)^l\left(\frac{-1}{\sinh  r} \frac{\diff}{\diff  r}\right)^{l+1}
\cos(qr) \\
N\kl &\equiv& \sqrt\frac2\pi |q| \left[ \prod_{n=1}^l
(q^2+n^2) \right]\mhalf \hspace{5em} (l>0)\eea
with $N_{k0} \equiv \sqrt{2/\pi} |q|$.
For $-1<q^2<0$, $\cos(qr)$ is replaced by $\cosh(|q|r)$.
The spectrum $\calp_f$ is defined by
\be
\langle \tilde f^*_{klm} \tilde f_{k'l'm'}\rangle =
 \frac{2\pi^2}{k|q^2|} \calp_f(k) \delta(k-k')
\delta_{ll'} \delta_{mm'}
\label{fullspec}
\ee
and the correlation function is
\be
\xi_f(  r)=\int^\infty_0 \frac{\diff k}{k}
\calp_f(k) \frac{\sin(q  r)}{q\sinh  r}
\label{correlation}
\ee

For $  r\gg 1$, the contribution to the correlation function from a mode with
$k^2\ll1$ is $\xi_f(  r)\propto \exp(-k^2  r)$
 Thus the correlation length, in units of the curvature
scale $a$, is of order $ k\mtwo$.
This is in contrast with the flat-space case, where
the contribution from a  mode with $k\ll 1$
gives a correlation length of order
$1/k$.

\bigskip
\begin{center}
{\large \bf The Grishchuk-Zeldovich effect}
\end{center}

\smallskip
\noindent
The cmb anisotropy is generally ascribed to a primeval curvature
perturbation (measured by comoving observers), conveniently taken
to be $\calr$ defined by
 \be 4(k^2+3) \calr\klm/a^2 =\delta \rt\klm \label{twei} \ee
In the limit $\Omega\to 1$,
 \be 4k^2 \calr\klm/a^2 =\delta \rt\klm \label{tweiflat} \ee
During matter domination
$\calr$ is constant
until $\Omega$ breaks away from 1.
For $l\lsim 30$ the
mean square multipole $C_l$ of the cmb anisotropy is given by
the Sachs-Wolfe approximation
\bea
C_l&=&
2\pi^2 \int^\infty_0 \frac{\diff k}{k}
\calp_\calr(k) I\kl^2 \label{openswsc} \\
|q| I\kl&=&\frac15 \Pi\kl(\eta_0) +\frac65 \int^{\eta_0}_0 \diff  r
\Pi\kl(  r)F'(\eta_0-  r)
\label{74}\\
F(\eta)&=&5\frac{\sinh^2\eta-3\eta\sinh\eta+4\cosh\eta-4}{(\cosh\eta-1)^3}
\eea
with $\eta=2(aH)\mone$ and $r=\eta_0$ the last scattering surface.
In this regime the COBE measurements give $l^2C_l\simeq
8\times 10^{-10}$. Within the observational uncertainties this is
consistent${}^4$ with a flat spectrum for all $0.1\leq\Omega_0\leq 1.0$,
of magnitude $\calp_\calr\sim 10^{-9}$ to $10^{-10}$. The corresponding
mean square curvature perturbation $\langle \calr^2 \rangle$ is of the
same order.

Now suppose that the spectrum $\calp_\calr$ rises sharply on some very
large scale $k\sub{VL}$ {\em but that the perturbation is still a typical
realization of a Gaussian random field} so that it can be discussed
using the above formalism. What is the effect on the cmb anisotropy?
For $\Omega_0=1$ this question was
asked and essentially
 answered by Grishchuk and Zeldovich${}^5$.
The large scale contribution can be taken to be
\be
\calp_\calr\su{VL}\simeq \delta(\ln k-\ln k\sub{VL})
\langle \calr^2 \rangle
\label{107}
\ee
In a sphere whose radius is of order the correlation length
$d\sub{VL}=a_0/k\sub{VL}$, $\calr$ is roughly
constant with typical value $\langle\calr^2\rangle\half$.
{}From \eq{tweiflat} it is of order $d\sub{VL}^2\delta  R^{(3)}$, which is a
dimensionless measure of the geometry distortion due to the
perturbation (recall that the background curvature $R^{(3)}$ vanishes
for $\Omega_0=1$). For instance it is roughly equal to the fractional
departure
from $4\pi d^2\sub{VL}$ of the sphere's area.
The maximal distortion, corresponding to
regions of positive curvature closing on themselves, is
$\langle \calr^2 \rangle\lsim 1$.
The quadrupole dominates for $d\sub{VL}\gg H_0\mone$, and
is given by
\be
C_2\su{VL}\sim (d\sub{VL} H_0)\mfour \langle \calr^2 \rangle
\ee
If the geometry distortion is maximal, then $d\sub{VL}H_0\sim
(C_2\su{VL})\quarter\gsim 10^2$. In words, the correlation length
is then more than two orders of magnitude bigger than the size of the
observable universe.

To generalize this analysis to
 $\Omega_0<1$ one needs to take the background spatial curvature into account,
and to note that the limit of large scales corresponds to
$k\to 0$, not $q\to 0$. This has not been done to date.
The only relevant publications of which we are aware either ignore the
spatial curvature${}^6$ or take use the $q\to0$  limit${}^{7,8}$.
Consider therefore \eq{107} with $k\sub{VL}\ll 1$,
and suppose that $\Omega_0$ is significantly below 1 so that
$a_0 H_0\sim 1$. In the absence of perturbations the geometry
distortion of the observable universe is $a_0^2 R^{(3)}
\sim 1$, and the addition distortion caused by the perturbation
is $\sim a_0^2 \delta R^{(3)}\sim \calr$. Thus, $\calr$ measures the
fractional change in the geometry distortion, and
since the relation between
$\delta R^{(3)}$ and $\calr$ is now scale independent this remains
true on larger scales. The maximal distortion,
corresponding to regions of space closing on themselves, is still
$\langle\calr^2\rangle\lsim 1$.

As $k\to0$, $\Pi_{k0}\to 1$,
but the other radial functions are proportional to
$k$. Defining $I_l\equiv \lim_{k\to0} I\kl/k$,
\be
C_l\su{VL} = I_l^2 k^{2}\sub{VL} \langle \calr ^2 \rangle
\label{gzlowom}
\ee
Since $I_l$ is roughly of order 1 for low multipoles, and also
$a_0\sim H_0\mone\sim 1$ we can write this
\be
C_l\su{VL}\sim (d\sub{VL} H_0) \langle\calr^2\rangle
\ee
The prefactor is not now raised to the fourth power as it is for
$\Omega_0=1$, so that for
maximal distortion $d\sub{VL}$ must now be ten orders of magnitude
bigger than the size of the observable universe!
There are two physical reasons for the difference.
One is that the correlation length is $a_0k\sub{VL}\mtwo$
instead of $a_0k\sub{VL}\mone$. The other
is that the presence of background curvature
allows the geometry distortion to be of order 1
in the observable universe, whereas before it was  at most of order
$k\sub{VL}^2\ll 1$.

In the case $\Omega_0=1$, the Grishchuk-Zeldovich effect contributes
only to the quadrupole, and is not seen in the data (ie., the quadrupole
is not anomalously high).
In the case $\Omega_0<1$ it contributes to all multipoles up to some
maximum, which is probably within the regime of validity of the
Sachs-Wolfe approximation. It would be worth evaluating the $l$ dependence
to see whether it is the same as the observed $C_l\propto l\mtwo$ for
some range of $\Omega_0$. If so the effect might be present, and one
could see whether this was so by looking at higher multipoles.
(A more bizarre possibility
would be that the effect persists even beyond the range of the
Sachs-Wolfe approximation, in which case a full calculation would be
necessary. The formalism is already in place, and
has already been used for the sub-curvature modes${}^{9,10,11}$.)

For ease of visualization we have used the concept of
the correlation length
$d\sub{VL}$, which presupposes that the perturbation continues to
be a typical realization of a Gaussian random field in a region
around us whose size is bigger than $d\sub{VL}$, and therefore much
bigger than the observable universe.
The effect is really
calculated on the hypothesis that the perturbation
is a typical realization of a
Gaussian random field
{\em within the observable universe}, and can be written
in terms of $k\sub{VL}$ without reference to a correlation length.
However, if the hypothesis is valid for $k$ down
to some minimum value, it is reasonable to suppose that it can indeed be
extended out to a region bigger than the corresponding correlation
length. Thus a positive detection of the
Grishchuk-Zeldovich effect would suggest that this is the case.
On the other hand a failure to detect the
effect, which seems more likely, will tell us essentially nothing!

Finally, let us ask whether one should expect the effect to be present
even below the level of detectability.
For the case $\Omega_0=1$ the
usual hypothesis is that the
curvature perturbation comes from a vacuum fluctuation of the inflaton
field, and in 1990 this was extended${}^{13}$
 to the case $\Omega_0<1$. To the extent that this is true
there are no super-curvature modes, which means that for
$\Omega_0$ appreciably less than 1 there is no Grishchuk-Zeldovich
effect. Like any hypothesis in physics this will be at best
an approximation, and it will fail above some large scale.
(As we just discussed, `scale' strictly means simply some large value of
$k\mone$, but one can probably think of it a also a large correlation
length.) However, the hypothesis that the curvature perturbation
in the observable universe is a typical realization of a homogeneous Gaussian
random field will also fail above some large scale, and
this might well be the same as the scale on which the vacuum fluctuation
hypothesis fails. If so, there will be no Grishchuk-Zeldovich effect.

It would be instructive to see how all this works in
bubble model${}^{13}$ of the $\Omega<1$ universe. According to this model
we inhabit the interior of the bubble extending far beyond the
observable universe. Within the bubble
the perturbation is well approximated by
a typical realization of a random Gaussian field, which
has only sub-curvature modes because it originates as a
vacuum fluctuation. As the boundary is approached the nature of the
perturbation changes and it no longer corresponds to a typical realization
of the random field. Thus one expects in this model the coincidence of
scales mentioned in the last paragraph, and no Grishchuk-Zeldovich
effect.

\bigskip
\begin{center}
{\Large \bf ACKNOWLEDGEMENTS}
\end{center}

\smallskip
\noindent
This work was started with the help of EU research grant
ERB3519PL920782(10835).
 One of us (DHL) thanks the Isaac Newton Institute for a visiting
Fellowship while the work was being completed,
and    Bruce
Allen,  Robert Caldwell,
 Misao Sasaki and Neil Turok
for useful discussions.

\bigskip
\begin{center}
{\large \bf REFERENCES}
\end{center}

\smallskip
\small
\noindent
1. YAGLOM, M. 1961 {\em in} Proceedings of the Fourth Berkeley Symposium
Volume II, J. Neyman, Ed. University of California Press, Berkeley.
\\
2. KREIN, M. G. 1949. Ukrain. Mat. Z. {\bf 1}, No. 1, 64;
{\em ibid} 1950 {\bf 2}, No. 1, 10.\\
3. LYTH, D. H. \& A. WOSZCZYNA. preprint astro-ph/9501044, submitted
to Phys. Rev. D.
\\
4.  KAMIONKOWSKI, M., D. N.  SPERGEL \& N.  SUGIYAMA. 1994.
Astrophys.  J.  {\bf 426}, L57;
GORSKI, K. M., H. RATRA, N. SUGIYAMA \& A. J.
BANDAY, preprint.\\
5. GRISHCHUK, L. P. \& Ya. B. ZELDOVICH. 1978.
Astron. Zh.  {\bf 55}, 209 [Sov. Astron. {\bf 22}, 125 (1978)].\\
6. TURNER, M. S.. 1991. Phys Rev D {\bf 44}, 12.\\
7. KAMIONKOWSKY, M.  \& D.  N SPERGEL. 1994.
 Astrophys.  J. {\bf 432}, 7.\\
8. KASHLINSKY, A., I.   TKACHEV \& J.  FRIEDMAN. 1994.
Phys. Rev. Lett., {\bf 73}, 1582.\\
9. SUGIYAMA, N. \& J.  Silk. 1994. Phys. Rev. Lett. {\bf 73}, 509.\\
10. KAMIONKOWSKI, M., D.  N.  SPERGEL \&  N.  SUGIYAMA. 1994.
Astrophys.  J.  {\bf 426}, L57.\\
11. GORSKI, K. M. , H. RATRA, N. SUGIYAMA \& A. J. BANDAY, preprint.\\
12. LYTH, D.  H.\& E.  D.
STEWART. 1990. Phys Lett. B {\bf 252}, 336.\\
13. COLEMAN, S. \& F. DE LUCCIA. 1980. Phys. Rev.
D  {\bf 21},  3305; GOTT, J. R.. 1982.  Nature {\bf 295} , 304;
GUTH, A.  H. \& E. J. WEINBERG. 1983.  Nucl. Phys. {\bf B212},
321; GOTT, J. R. \& T. S. STATLER. 1984.  Phys. Lett. {\bf B136},
157; SASAKI,  M., T. TANAKA, K. YAMAMOTO \& J. YOKOYAMA. 1993.
Phys. Lett. B  {\bf 317}, 510;
SASAKI, M., T. TANAKA, K. YAMAMOTO \& J. YOKOYAMA. 1993.
Prog. Theor. Phys. {\bf 90}, 1019;
BUCHER, M., A. GOLDHABER \& N. TUROK. 1994.  preprint;
TANAKA, T. \& M. SASAKI. 1994. two preprints;
YAMAMOTO, K., T. TANAKA, \& M. SASAKI. 1994.  preprint.

\end{document}